\documentclass[journal]{IEEEtran}

\usepackage{amsmath,amsthm,amstext,amsfonts,amssymb,mathrsfs}
\usepackage{graphicx,caption,color,algorithm,algorithmic} 
\usepackage{xcolor}
\usepackage{makecell, hhline}
\usepackage{subfig} 
\usepackage{url}

\theoremstyle{remark}

\title{Deep Reinforcement Learning-Based \\ Robust Protection in DER-Rich Distribution Grids}

\author{Dongqi~Wu,~\IEEEmembership{Student~Member,~IEEE,}
	Dileep~Kalathil,~\IEEEmembership{Senior~Member,~IEEE,} Miroslav~Begovic,~\IEEEmembership{Fellow,~IEEE,}
	and~Le~Xie,~\IEEEmembership{Senior~Member,~IEEE}
	\thanks{Authors are with the Department
		of Electrical and Computer Engineering, Texas A\&M University, College Station,
		TX-77843, USA. e-mail: \{dqwu, dileep.kalathil, begovic, le.xie\}@tamu.edu. Corresponding Author: Le Xie.}}

\begin{document}

\maketitle

\begin{abstract}
This paper introduces the concept of Deep Reinforcement Learning based architecture for  protective relay design in power distribution systems with many distributed energy resources (DERs). The performance of widely-used  overcurrent protection scheme is hindered by the presence of  distributed generation, power electronic interfaced devices and fault impedance. In this paper, a  reinforcement learning-based approach is proposed to design and implement protective relays in the distribution grid. The particular algorithm used is an Long Short-Term Memory (LSTM) enhanced deep neural network that is highly accurate, communication-free and easy to implement. The proposed relay design is tested in OpenDSS simulation on the IEEE 34-node test feeder and demonstrated much more superior performance over traditional overcurrent protection from the aspect of failure rate, robustness and response speed.
\end{abstract}


\IEEEpeerreviewmaketitle

\section{Introduction}

\IEEEPARstart{T}{his} paper proposes and conceptually tests a novel Deep Reinforcement Learning (Deep RL) based approach for protective relay control design in distribution grids. Recent developments in photovoltaic (PV) and power electronics  technology have led to an increase of penetration of distributed energy resources (DER) in distribution grids. DERs, especially solar PVs, can provide a number of benefits to the power system operation efficiency such as peak load reduction and improved power quality \cite{DGBenefits}. However, DER and emerging grid edge-level devices are increasing the complexity of the interactions between end users and distribution grid operators substantially, such as low or non-existent system inertia, islanded operation and load-side voltage security. These additional complexities pose significant challenges for the operation and protection of the distribution grid.

Protective relays are the safeguards of distribution systems. The role of protective relays is to protect the grid from sustaining faults by disconnecting the smallest practically available faulty segment from the rest of the grid. During the operation, a relay monitors the power grid and looks for patterns that are associated with faults. Typical measurements include current (over-current and differential relay), voltage and current (distance relay), frequency or electromagnetic wave from transients (traveling-wave relay). In power distribution systems,  time delayed, coordinated overcurrent relays are most commonly used since many other methods are impractical due to cost, infrastructure and grid topology limitations.

\begin{figure}[t]
\centering
\includegraphics[scale=0.4]{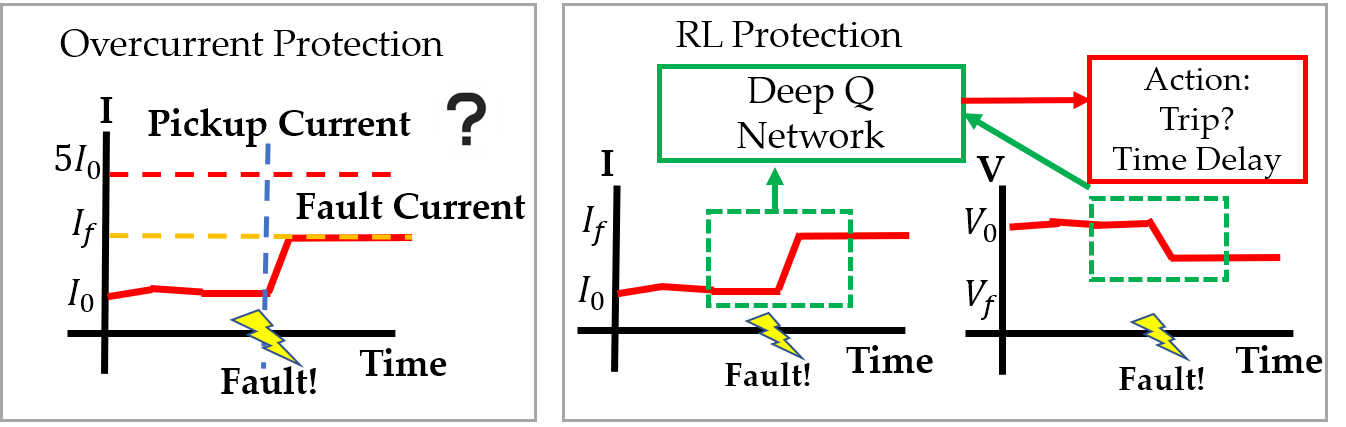}
\caption{Concept of Overcurrent and RL Protection}
\label{fig:concept}
\vspace{-0.3cm}
\end{figure}

However, it is very difficult for overcurrent relays to accommodate the vastly different operational conditions in real distribution grids. For feeder recloser relays, the presence of DER within the feeder can reduce the fault current measured at the recloser and make faults harder to detect. The fault current contribution from DERs to the fault will also make the fault current observed at the fuse higher than the current at the recloser, making coordination based on inverse-time curves difficult \cite{DGProtectionSurvey} \cite{34busovercurrent1}. Moreover, even in current distribution grids, factors like fault impedance and load profile change are not taken into account in the traditional overcurrent protection design, resulting in problems such as failing to detect faults near the end of a feeder, a.k.a. under-reaching. Fig. \ref{fig:concept} shows a conceptual comparison between threshold-based overcurrent protection and our proposed RL protection. Where overcurrent relays may be affected by low fault current magnitude, our method does not suffer the same limitation as it detects faults using the waveform patterns in measurements.

Traditional protective relays are also designed to function under two crucial assumptions: (i) power flow is unidirectional from the substation towards the end users, and (ii) the difference between operating conditions (currents and voltages) between normal and faulted conditions are measurable and significant. With the increasing popularity and penetration of DER and grid-edge devices, both assumptions will likely be rendered invalid \cite{DGProtectionSurvey}.  For example, in the simple circuit shown in Fig. \ref{fig:radial1}, there is a distributed generator feeding power into the grid at bus B. Under conditions where the net power absorption of the loads at bus B and C is low, or the output of the distributed generator at B is having a high peak, the power flow direction in the line between A and B will be from B to A, which violates assumption (i). For a fault to the right of bus B as indicated in the figure, the fault current contribution of the distributed generator could decrease the magnitude of fault current measured at the recloser bus A to the range of peak load current under normal conditions and potentially violating assumption (ii). In fact, reliable protection is becoming an Achilles' hell that limits the growth of DER integration for future grids.

\begin{table}
    \begin{center}
        \caption{Fault current level under various DER fraction of total load and fault impedance}
        \begin{tabular}{|c|c|c|c|c|}
        \hline
        DER \% \textbackslash{} $Z_f$($\Omega$)  & 0.01 & 0.1  & 1    & 5    \\ \hline
        0                                      & 3.841 & 3.676 & 2.197 & 1.283 \\ \hline
        25                                     & 3.875 & 3.653 & 1.995 & 1.074 \\ \hline
        50                                     & 3.913 & 3.629 & 1.794 & 0.877 \\ \hline
        75                                     & 3.956 & 3.603 & 1.596 & 0.697    \\ \hline
        \end{tabular}
        \label{table:simple}
    \end{center}
    \vspace{-0.3cm}
\end{table}

We further demonstrate the challenge posed by DER to protection using the simplest numerical example. Consider the same radial feeder in Fig\ref{fig:radial1}. This circuit has only one load at bus C, a distributed generator is placed at bus B. This generator is modeled as constant for the purpose of this example with current limit to roughly mimic an inverter based supply. The line parameters are adopted from the IEEE 4 bus feeder system. The rated current is calculated without the distributed generator and all power are supplied by the substation through the transmission grid. Under this simple illustrative setting, we vary the capacity of the distributed generator as a percentage of the total load, add a single-phase fault with various level of fault impedance and record the current at bus A. The ratio between fault current and load current is listed in Table \ref{table:simple}.

It can be clearly seen that the presence of distributed generator and fault impedance can greatly reduce the magnitude of fault current. Usually, for overcurrent relays, the fault current needs to be at least 2 to 3 times higher than the normal operation current under maximum load level to detect faults reliably. Under the impact of DER or fault impedance, or both, the fault current magnitude can be too low to detect for overcurrent relays. In contrast, for this simple example, we will demonstrate that our proposed RL relay is able to successfully detect faults under all scenarios above.

\vspace{-0.3cm}
\subsection{Literature Review} 

There are many studies on  improving the performance of protective relays. Most of them focus on improving  the performance  of  commonly used overcurrent relays by better fault detection \cite{Lit01} and coordination \cite{Lit02}. Neural networks have been used in setting the parameters of overcurrent relays \cite{Lit03}. Support Vector Machine (SVM) can be trained to distinguish the normal and fault conditions directly \cite{Lit04}\cite{Lit05}. A recent work \cite{Lit06} uses tabular Q-learning to find the optimal setting for overcurrent relays. Most proposed methods are still confined within the framework of inverse-time overcurrent protection, which is considered not enough for the future distribution grid with high DER and EV penetration \cite{DGProtectionSurvey}.

Reinforcement Learning (RL) is a  branch of machine learning that addresses the problem of learning optimal control policies for unknown  dynamical systems. RL algorithms using deep neural networks \cite{DQN}, known as  \text{Deep RL} algorithms, have made significant achievements  in the past few years in areas like  robotics, games, and autonomous driving  \cite{DeepRLSurvey}. RL has also been applied to various power system control problems including voltage regulation \cite{RLGarcia}, frequency regulation \cite{RLFreqReg}, market operation \cite{RLMarket}, power quality control \cite{RLPowerQuality} and generator control \cite{RLAGC}. Our previous paper \cite{RelayCDC} was the first work to use deep RL for power system protection. A comprehensive survey of RL applications in power system is detailed in a recent review paper \cite{GlavicRLSurvey}.

%
%
%

\vspace{-0.3cm}
\subsection{Main Contributions}
This paper introduces a novel deep RL based framework design robust protective relays in distribution grids with many DERs. Key contributions are suggested as follows:


\begin{itemize}
	\item To our best knowledge, we have for the first time formulate the design of protective relay in radial distributed systems as a nested RL problem.  
	\item Based on the new formulation, we proposed a novel Long-Short-Term-Memory (LSTM)-enhanced RL algorithm that leads to much more reliable and accurate coordination of protective relays as compared to conventional inverse time over-current relays. 
	\item We develop a fully automated software interface that complies with OpenAI Gym that is readily available to integrate state-of-the-art machine learning packages and commercial grade power system simulators. 
\end{itemize}
This work is substantial expansion upon \cite{RelayCDC}. Major improvements includes detailed formulation to three-phase unbalanced scenarios, an LSTM enhanced network model, and diverse but realistic test scenarios.
The rest of this paper is organized as follows: Section \ref{sec:formulation} formulates the protection problem using the framework of RL. Section \ref{sec:nestedRL} introduces our nested reinforcement learning algorithm for relay control problem. Section \ref{sec:experiment-setting}  presents the simulation environment and the test-bed used in training and evaluation,  Section \ref{sec:simulations} analyzes and discusses the simulation results. Section \ref{sec:conclusion} summarizes and concludes the paper.
\vspace{-0.1cm}
\section{Problem Formulation}
\label{sec:formulation}


%
\subsection{Relay Operation}

\begin{figure}[t]
\centering
\includegraphics[scale=0.5]{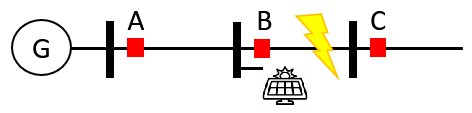}
\caption{Diagram of a Simple Distribution Feeder}
\label{fig:radial1}
\end{figure}

\begin{figure}[t]
\centering
\includegraphics[scale=0.6]{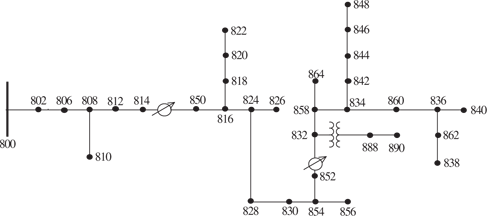}
\caption{IEEE 34 node test feeder}
\label{fig:IEEE34}
\vspace{-0.3cm}
\end{figure}

We first illustrate the ideal operation of relays using a simple distribution feeder as given in Fig. \ref{fig:radial1}. There are 3 relays and breakers located at each bus of the distribution line.  Each relay is located to the right of a bus. Each relay needs to protect the segment that is between its location and the buses and loads located downstream. The protection of inverter-based generators is not considered to be part of the functions of the relays. Thus, it effectively isolates inverters from the rest of the network. Each relay except relay C is also required  to provide backup protection for its downstream neighbor: when its neighbor fails to operate, it needs to trip the line and clears the fault after a reasonably short wait time. For example, in Fig. \ref{fig:radial1}, if a fault occurs at the point where the broken arrow indicator is, relay B is the main relay protecting this segment and it should trip without any intentional delay. If relay B fails to trip, relay A, which provides backup for relay B, needs to trip instead, after a sufficiently short delay of the order of a fraction of a second, to allow primary protection to trip first, if appropriate. The time delay between the fault occurrence and relay tripping should be as short as possible for primary relays, while backup relays should react slower to ensure that they are  triggered only when the corresponding primary protection is not working. The coordination time between primary and backup protection should be short enough to allow the aggregate coordinated time response at the relay closest to the substation to effectively trip  the fault currents.

\vspace{-0.4cm}
\subsection{Reinforcement Learning and Modeling of RL Relay}
In RL formulations, a control problem is modelled as an active interaction between the controller, a.k.a. \textit{Agent}, and the system, or environment, to be controlled. The system is represented by a Markov Chain whose state evolves based on a deterministic or stochastic transition kernel as well as the actions of the agent. The agent observes the state of the system and give control actions based on its \textit{policy}. Each action is assigned a \textit{reward} that is based on the effect of the action and the resulting state transition. In the process of solving an RL problem, the agent learns a control policy that gives the most optimal action corresponding to each observed system state in order to maximize total expected reward. Unlike traditional control problems in which the controller is derived from analytical development of an accurate model of the system, or plant, an RL agent learns its optimal policy through extensive observation under perturbations of the system state. The RL agent typically assume no prior knowledge about the system model at the beginning of learning, it then gather experience about the system state transition and reward by attempting different actions under different states. After enough experience is collected, the agent will be able to choose the actions that results in the highest long-term reward based on the observations it receives.

\begin{figure*}[t]
	\centering
	\includegraphics[scale=0.35]{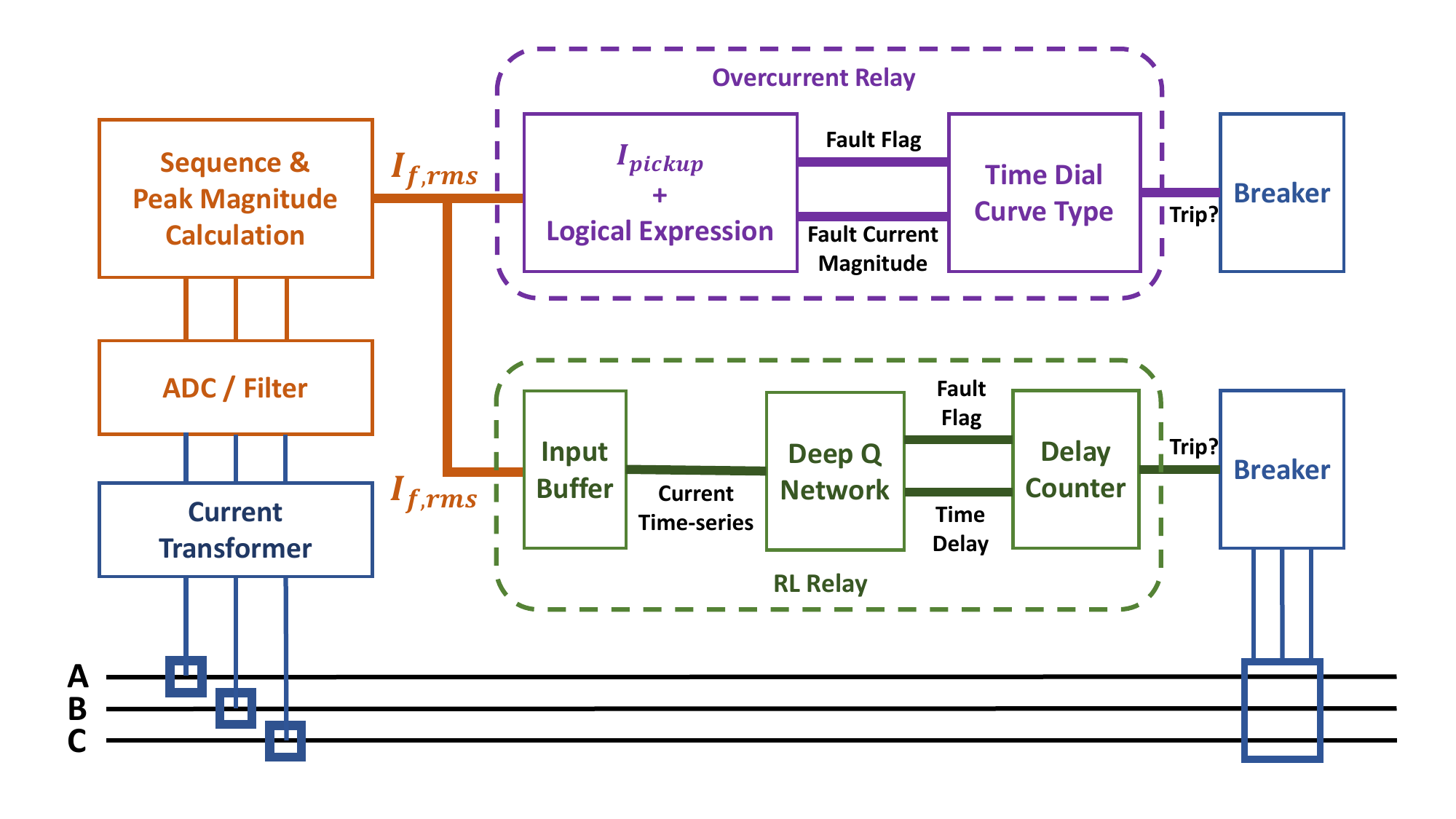}
	\caption{Conceptual Comparison Diagram between Overcurrent and RL relay}
	\label{fig:flowchart}
\end{figure*}


We propose a RL based relay control strategy that is adaptive and robust even in DER-rich distribution grids. The RL relay can take the same or more measurements available to traditional overcurrent relays and output a tripping signal with a time delay for coordination without communication when it detects a fault. Unlike the threshold-based fault detection logic that are designed from a finite set of scenarios and strong assumptions, the RL relay can learn a policy that is based on not only the instantaneous post-fault current, but also pre-fault condition and system dynamics from the transient response. A flow diagram for comparison of the concept of standard overcurrent relay and our RL relay design is shown in Fig. \ref{fig:flowchart}. Under the RL framework, the RL relay is an agent that monitors the grid using local current measurements, assess the real-time condition using a trained policy and output trip signals accordingly. During the training of the policy, the agent explores a large number of current measurement around the time of fault events and learn the pattern associated with those events as well as the correct response. To facilitate the training process, a synthetic model is needed to produce the large amount of training data. This is achieved by building a simulation environment that can generate random fault scenarios and adjust according to the agent's actions. The minutiae of implementation are discussed in section 3B.

This formulation using reinforcement learning, compared to other data driven or machine learning based technologies, has several exclusive advantages that are especially appropriate for modeling the decision making of protective relays. First, most methods such as support vector machine or artificial neural network take a supervised learning approach, which attempts to develop a best classifier to distinguish normal and fault conditions from the training data-set. These methods require all training data to be properly labeled in advance which, in this case, means the optimal relay action and associated time delay for each series of current measurement in the training data. Although supervised learning methods work reasonably well in identifying faults, they cannot determine the optimal time delay for accurate coordination because it varies among scenarios and is difficult to quantify and justify at the time of training. In contrast, in RL explicit labeling of the training data is not required. The agent only need to be told that if the action it performs is desirable, which can be easily determined based on the status of the circuit. In short, supervised learning is an instructive process where the teacher need give step-by-step instructions to tell the agent exactly what to do during training, while RL is an evaluative process where the trainer only need to provide an evaluation of the actions taken by the agent based on the final outcome. The latter is more suitable in a dynamic control problem such as in the electric grid where typically a well-formulated objective need to be achieved. Second, it can be difficult to incorporate the underlying model of the system in other data-driven techniques. Consecutive measurement taken at the current/potential transformer by relays are determined by the network model, which is usually very complex or computationally intensive to obtain and utilize. However, an underlying system model is included in the formulation of RL, and during the learning process it learns the patterns in state transition and evolution in order to make a series of consecutive actions to achieve a desirable final objective.


\vspace{-0.2cm}
\subsection{Mathematical Formulation of Markov Decision Processes and Reinforcement Learning}
Next, we will proceed to give a brief review of the basic concepts of Markov Decision Process (MDP) and RL and then present a mathematical formulation of the protective relay problem. This formalism will be expanded later to formulate the optimal control for relay protection problem under the framework of multi-agent reinforcement learning. A concise but more comprehensive introduction of MDP, Dynamic Programming (DP) and RL could be found at \cite{RLtutorial}.

\emph{Markov Decision Processes} (MDP) is a mathematical framework for stochastic control problems. This framework models a control problem as a sequential decision making problem where the environment varies partly random and partly based on the control actions. An MDP is modeled as a $tuple$ with 5 elements: $(\mathcal{S}, \mathcal{A}, R, P, \gamma)$ in which $\mathcal{S}$ is the state space, $\mathcal{A}$ is the action space. $P = (P(\cdot|s, a), (s, a) \in \mathcal{S} \times \mathcal{A})$ is the transition probabilities that corresponds to probability of transitioning to state $s'$ from state $s$ as a result of action $a$. $R : \mathcal{S} \times \mathcal{A} \rightarrow \mathbb{R}$ is the reward function and $\gamma \in [0, 1)$ the discount factor.   Under the RL framework, for a protective relay, its state space $S$ will include all information available for observation (e.g. voltage, current or frequency); its action space $A$ will include all possible breaker operation; $P$ will be determined by the model of the distribution grid where the relay is deployed; reward $R$ and discount factor $\gamma$ will be chosen before training to promote desirable operations. 

Under each state, the action of an agent is given by its policy $\pi : \mathcal{S} \rightarrow \mathcal{A}$ .A policy is usually evaluated using the \textit{value function}, $V_{\pi}$:
\begin{align*}
&V_{\pi}(s) = \mathbb{E}[\sum^{\infty}_{t=0} \gamma^{t} R_{t}   | s_{0} =s ],
\end{align*}
where $R_{t} = R(s_{t}, \pi(s_{t}))$ and $s_{t+1} \sim P(s_{t}, a_{t})$. 
The $optimal$ $value$ $function$, $V^{*}$ is the value function of the optimal policy that gives the highest value function: $V^{*}(s) = \max_{\pi} ~V_{\pi}(s)$. The optimal policy $\pi^{*}$ can also be obtained from $V^{*}$ using the Bellman equation:
\begin{align*}
\pi^{*}(s) = \arg \max_{a \in \mathcal{A}} ~(R(s, a) + \gamma \sum_{s' \in \mathcal{S}} P(s'|s, a) V^{*}(s') ).
\end{align*} 

The Q-value function of a policy $\pi$, $Q_{\pi}$, is defined as $
Q_{\pi}(s, a) = \mathbb{E} [\sum^{\infty}_{t=0} \gamma^{t} R_{t}  | s_{0} =s, a_{0} = a ]$. The Optimal Q-value function $Q^{*}$ is also defined similarly, $Q^{*}(s, a) = \max_{\pi} Q_{\pi}(s, a)$. The optimal policy could be directly obtained using the optimal Q-value function as $\pi^{*}(s) = \arg \max_{a \in \mathcal{A}} ~ Q^{*}(s, a)$ 
Specifically, the optimal Q-value for tripping the breaker will be the highest when a fault is present on the feeder, which improves the reliability of relays; the optimal Q-value for staying closed will be the highest when the system is under normal conditions, which enhances the dependability of relays.

For MDP formulations, the optimal value/Q-value function or the optimal control policy can be computed using dynamic programming  methods \cite{sutton}. These explicit methods requires the full transition probability matrix $P$ for all possible state variable combinations. However, in realistic problems the exact system model is usually difficult to obtain. Specifically, in the protective relay problem, the transition probability  represents all possible stochastic variations in the phasor voltage and current in the network caused by uncertainties in load profile and DER generation. Fault parameters such as fault impedance and location are also rendering difficult accurate determination of that probability because some parameters of the model can vary widely from one fault to another. Learning based methods are generally more appropriate for such problems with high level of uncertainties.

\textit{Reinforcement learning}  is a method for learning the optimal policy for an MDP when a explicit model is not available. In RL, the optimal policy is learned through sequential observations and interactions with the system. In \textit{Q-learning}, which is the most commonly used algorithm for RL, the optimal Q-value function is learned from a sequence of interactions $\left(s_t, a_t, R_{t}, s_{t+1} \right)$. Specifically, the Q-value function $Q_{t}$ is updated at each time step $t$ as:  
\begin{align}
Q_{t+1}(s_{t},a_{t}) &= Q_t(s_{t},a_{t}) + \alpha_{t} [ R_{t} +  \nonumber \\ 
&\hspace{0.5cm} \gamma \max_{b\in A} Q_{t}(s_{t+1},b) - Q_t(s_{t},a_{t})]
\end{align}
where $\alpha_{t}$ is the  step size, or learning rate, that determines how fast newly collected information gets incorporated to the existing model. With an appropriately chosen $\alpha$, $Q_{t}$ will converge to the optimal Q-function $Q^{*}$ after each state-action pair has been visited sufficiently often \cite{sutton}.

The standard  Q-learning algorithm as described cannot be directly used in problems with continuous state/action space. For continuous problems, a deep neural network is usually used as a replacement for an explicit Q-function: $Q(s,a) \approx Q_{n}(s,a)$ and $n$ represents parameters of the neural network. The neural network for Q-learning is usually called Deep-Q-Networks (DQN).The ability of neural networks to \textit{approximate any function} using only input-output samples has enabled tremendous success in many reinforcement learning problems in different fields.

For each state-action pair $\left(s_t, a_t, R_{t}, s_{t+1} \right)$, the parameters of the DQN can be updated using stochastic gradient descent:  
\begin{align}
\label{eqn:sgd}
n &= n + \alpha \nabla Q_n(s_t,a_t) \nonumber  \\
&\hspace{1cm} (R_t + \gamma \max_{b\in A} Q_n(s_{t+1},b) - Q_n(s_t,a_t))
\end{align}

Q-learning with neural network can be improved by implementing various upgrade techniques. \textit{Experience replay} is used as a buffer to store a batch of observations and shuffle them before each gradient upgrade. This can help to avoid the bias introduced by the temporal correlation among observations obtained in a sequence. \textit{Target network} is a separate neural network model only used to temporarily fix the gradient descent target for several steps to avoid potential instability caused by chasing a moving target. 

\begin{figure}[t]
	\centering
	\includegraphics[scale=0.21]{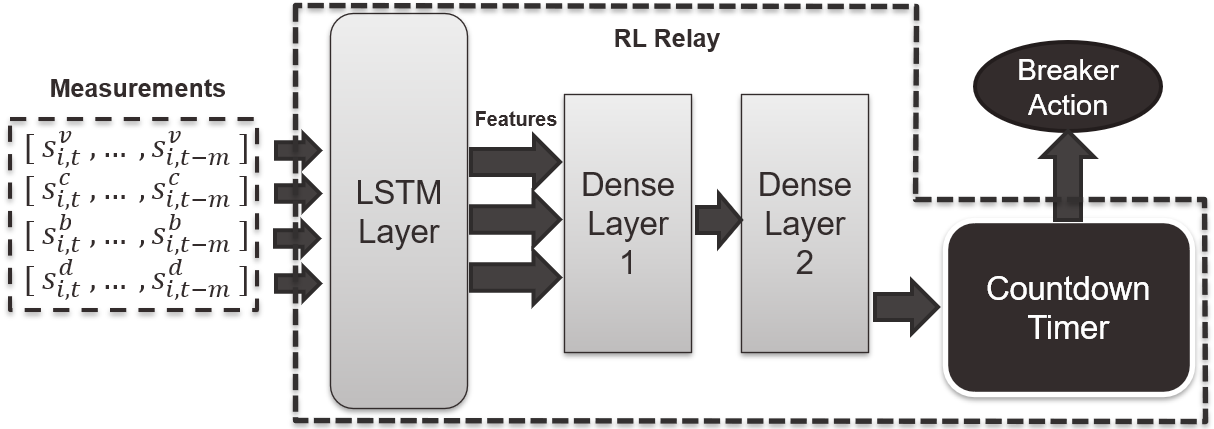}
	\caption{Model Structure of an RL Relay}
	\label{fig:struct}
\end{figure}

An Long-Short-Term-Memory(LSTM) layer \cite{lstm} is used before the fully-connected layers to extract features from time series inputs. Using LSTM with deep reinforcement learning \cite{lstmrl} has received increasing attention in recent years in time-correlated control problems. LSTM has a unique advantage over other non-recurrent neural network models, that is, the ability to remember what has happened in the past. Each LSTM cell has a internal state that can be either kept/changed/forgot for every observation it receives. This feature is particularly useful for assessing the current state of power systems, as it is able to adapt to the change of states incurred by other disturbances that does not need protection to operate (e.g. daily load curve, renewable generation profile, etc.). Our algorithm is based upon the combination of deep neural networks, experience replay, target network and LSTM feature extraction as illustrated by the flowchart in Fig. \ref{fig:struct}.

%
%
%


\vspace{-0.3cm}
\subsection{Protective Relay Control as an RL Problem}

We formulate the distribution system transient process as an MDP  {environment} and model the relays as RL {agents}.
For consistency with the current protection infrastructure, each relay is set to only observe its local current measurements $(s^{c}_{i,t})$,  although if additional information(voltage, frequency, etc.) is added in to the state space the RL relay would easily accommodate them without changing the formulation and potentially achieve even better performance. Each relay also knows the status of the local current breaker circuits, i.e., if it is open or closed  $(s^{b}_{i,t})$. Each relay also has a local counter that ensures the necessary time delay in its operation as a backup relay  $(s^{d}_{i,t})$.  These variables constitute the state $s_{i,t} = (s^{c}_{i,t}, s^{b}_{i,t}, s^{d}_{i,t})$ of each relay $i$ at time $t$. Table \ref{table:state_space} summarizes this state space representation.  Note that the each state also uses the past $m$ measurements to form a timeseries of measurement with length $m+1$. An appropriate combination of sampling rate and length of the timeseries allow one to deal with some classes of transients that cannot be identified from phasor measurements (such as inverter controls and limiters) in order to determine post-transient state.

Relay should operate after faults occur. However, since each relay is able to observe only its local state and no communication is assumed between the relays, some implicit coordination between relays is necessary. In traditional overcurrent protection scheme, the coordination is achieved using an inverse-time curve that adds a time delay between the detection of fault and actual breaker operation, based on the variation among fault current magnitudes at different locations on the feeder. However, fault current magnitudes can be unpredictable across different scenarios, especially with DER and smart edge-devices. We propose another approach (that is also amenable to RL) as follows. Instead of tripping the breaker instantaneously, it controls a countdown timer to indirectly operate the breaker. If a fault is detected, the relay can set the counter to a value such that the breaker trip after a certain time delay. The counter could be cancelled prematurely if the fault is cleared by another protective device. The action of each relay $i$ at time $t$, $a_{i,t}$ is summarized in table \ref{table:action_space}.

The reward given to each relay is a measure of success for its most recent action. A positive reward is given to an RL relay if: i) it remains closed during normal conditions, ii) it trips the breaker after a fault in the downstream circuit where it is the closest protection device, or when other closer protections fail to operate. A negative reward is given if: i) tripping the breaker when there is no fault, or the fault is outside of its assigned region; 2) fail to trip the breaker when a fault is present in its assigned region. The magnitude of the rewards are designed to implicitly signify relative importance of false positives (lack of dependability) and false negatives (lack of reliability). The reward function for each relay is shown in Table \ref{table:reward}.

The transition probability in a distribution feeder with multiple RL relays relates the change in power flow states to the measurement and operation of RL relays. Formally, let the global state at time $t$, $\bar{s}_{t} = (s_{1,t}, s_{2, t}, \ldots, s_{n,t})$, denote all nodal voltage and branch current in the system; let the combined action at time $t$, $\bar{a}_{t} = (a_{1,t}, a_{2, t}, \ldots, a_{n,t})$, denote the action of every RL relay in the system. Then, the state of the system $\bar{s}_{t}$ evolves stochastically based on $\bar{a}_{t}$ plus the variation in load profile, DER output and circuit connectivity. Note that the global state evolution cannot be described by local transition probabilities of individual relays because the action of any relay can affect the states of other relays. The global system dynamics is represented by the transition probability  $ \bar{P}(\bar{s}_{t+1} | \bar{s}_{t}, \bar{a}_{t})$. 


The goal in the multi-agent RL formulation is to achieve a global optimum which maximizes the expected sum of reward received by all relays, using only local control laws $\pi_{i}$ on local observations $s_{i,t}$: $\max_{(\pi_{i})^{n}_{i=1}} ~ \mathbb{E}[\sum^{\infty}_{t=0} \gamma^{t} \bar{R}_{t} ],~ a_{i,t} = \pi_{i}(s_{i,t})$. Local policies $\pi_{i}$ needs to be computed individually as a centralized policy would not be possible due to lack of communication. 


\begin{table}
	\begin{center}
		\caption{Relay State  Space}
		\label{table:state_space}
		\begin{tabular}{|c||c|}
			\hline
			State & Description\\
			\hline
			$s^{c}_{i,t}$ & Local current measurements of past $m$ timesteps\\
			\hline
			$s^{b}_{i,t}$ & Status of breaker (open (0) or closed  (1))\\
			\hline
			$s^{d}_{i,t}$ & Value of the countdown timer\\
			\hline
		\end{tabular}
		\vspace{-0.4cm}
	\end{center}
	
\end{table}

\begin{table}
	\begin{center}
		\caption{Relay Action Space}
		\label{table:action_space}
		\begin{tabular}{|c||c|}
			\hline
			Action & Description\\
			\hline
			$a_{\text{set}}$ & Set the counter to value to an integer between 1 and 9\\
			\hline
			$a_{d}$ & Decrease the value the counter by one\\
			\hline
			$a_{\text{reset}}$ & Stop and reset the counter \\
			\hline
		\end{tabular}
		\vspace{-0.4cm}
	\end{center}
	
\end{table}

\begin{table}[t]
	\begin{center}
		\caption{Reward for Different  Operations}
		\label{table:reward}
		\begin{tabular}{|c||c|}
			\hline
			Reward & Condition\\
			\hline
			Large Positive & \makecell{Tripping when a fault is present\\ in its assigned protection region}\\
			\hline
			Large Negative & \makecell{Tripping when there is no fault \\ or the fault is outside its assigned region}\\
			\hline
			Small Positive & \makecell{Stay closed when there is no fault \\ or the fault is outside its assigned region} \\
			\hline
			Small Negative & \makecell{Stay closed when a fault is present \\ in its assigned protection region} \\
			\hline
		\end{tabular}
	\end{center}

\end{table}

\vspace{-0.2cm}

\section{Nested Reinforcement Learning for  Control of  Protective Relays}
\label{sec:nestedRL}

In many distribution feeders there would be multiple protection devices coordinating with each other to provide extra security. However, obtaining the policies for a network of distributed RL relays operating in the same system could be difficult because, 1) Normal RL methods require the environment to appear stationary to the agent; 2) The whole system state in a power grid is not observable using measurements collected from only one location. Multi-Agent-RL(MARL) \cite{MARL} problems are often untrackable and the performance of available algorithm is generally not reliable. 

We proposed a \textit{Nested Reinforcement Learning} algorithm that cleverly takes advantage of the radial structure of  distribution systems to simplify the otherwise difficult MARL problem. In radial distribution systems, the dependency between the operation of coordinating relays is uni-directional, i.e., only upstream relays need to provide backup for a downstream relay but not vice-versa. Also, the last relay at the load side does not need to coordinate with others. In our nested RL algorithm, we start the RL training from the the most remote relay from the distribution transformer whose ideal operation is not affected by the operation of other relays, thus can be trained using a single-agent algorithm. Then, we can fix the trained policy for this last relay and train the relays at one-level closer to the substation that need to provide backup for the last relay. Since the policy of the furthest relay is fixed, it appears like a part of the stationary environment to its upstream neighbors which can learn to accommodate its operation. This process can be repeated for all the relays upstream to the substation. This method is analogous to how the coordination of time-delayed overcurrent relays is performed. The order of training can be determined by network tracing using a \textit{post-order depth-first} tree traversal with the substation being the root.  This nested training approach which exploits the nested structure of the underlying physical system allows us to overcome the non-stationarity in generic multi-agent RL settings. Our nested RL algorithm for training a system with $n$ RL relays is formally presented below.

\begin{algorithm}[ht]
	\caption{Nested Reinforcement Learning Algorithm}
	\label{algo:marl}
	\begin{algorithmic}
		\STATE Initialize DQN of each relay $i$ with random weights 
		\STATE Sort all relays based on system topology
		\FOR {relay $i = 1$ to $n$} 
		\FOR {episode $k = 1$ to $K$}  
		\STATE Initialize simulation with random system parameters
		\FOR {time step $t = 1$ to $T$} 
		\STATE Observe the state  $s_{i, t}$  for all relays
		\FOR {relay  $j = 1$ to $i$ (Trained Relays)}
		\STATE Select action using the trained policy as:
		\STATE $a_{j,t} = \arg \max_{a} Q_{n^{*}_{j}}(s_{j,t}, a)$
		\ENDFOR 
		\FOR {relay  $j = i+1$ to $n$}
		\STATE Select do nothing action, $a_{j,t} = 0$
		\ENDFOR
		\STATE With probability $\epsilon$ select a random action $a_{i,t}$, 
		otherwise select the action with the highest $Q$ value: $a_{i,t} = \arg \max_{a} Q_{n_{i}}(s_{i,t}, a)$
		\STATE Observe  reward $R_{i,t}$ and next state $s_{i,t+1}$
		\STATE Store $(s_{i,t}, a_{i,t}, R_{i,t}, s_{i,t+1})$ in the replay \\ buffer of relay $i$
		\STATE Sample a batch of past transitions from replay buffer and update the DQN parameter $w_{i}$
		\ENDFOR
		\ENDFOR
		\ENDFOR
	\end{algorithmic} 
\end{algorithm}

\section{Experiment Environment and Test Cases}
\label{sec:experiment-setting}

In this section, we  describe the simulation environment, test system modelling and experiment design. 

\vspace{-0.3cm}
\subsection{Simulation Environment}
The simulation environment is built by packing the OpenDSS APIs in a Python class inherited from the OpenAI Gym  \cite{gym} to improve accessibility. We note that this setting can  potentially be used in a number of other   research problems  addressing the distribution systems operation using machine learning. The RL algorithm is programmed in Python using open-source machine learning packages Tensorflow \cite{tensorflow}. The hyper-parameters of the DQN for each relay are selected through random search are are listed in Table \ref{tab_DQN} to serve as a starting point for potential replications of the works.

\begin{table}
\begin{center}
	\caption{DQN Hyper-parameters}
	\label{tab_DQN}
	\begin{tabular}{|c||c|}
		\hline
		Hyper-parameter & Value\\
		\hline
		 LSTM Cell Number & 70\\
		\hline 
	    Hidden Layers & 256/128\\
		\hline
		Activation & ReLU/ReLU/Linear \\
		\hline
		\makecell{Target Network Update Rate} & 0.005\\
		\hline
		Optimizer and Learning Rate & Adam, 0.0001 \\
		\hline
	\end{tabular}
\end{center}
\end{table} 

\vspace{-0.2cm}
\subsection{Test System Modeling}
We choose the IEEE 34-bus test feeder to test the performance of RL based recloser relay control. The test cases are replicated in  OpenDSS using the same parameters provided in IEEE publications \cite{IEEEcases}. Overall, OpenDSS power flow result and IEEE results agree closely, while the difference is mainly caused by aggregating distributed loads in a dummy bus at the midpoint of each branch. The percentage difference of node voltages between the OpenDSS simulations and IEEE published values are listed in Table \ref{pf_comparison}. 

 The RL recloser relay will be placed at the substation (bus 800), its only task is to respond to faults as quickly as possible. In distribution systems, it is common for large and long feeders to have additional reclosers in the middle of the feeder for additional security. For persisting faults that cannot be cleared by reclosing, the recloser needs to be locked open. In such cases, it is preferred for the closest protection device to operate to reduce the amount of load being disconnected and mitigate the damage. In the following multi-agent coordination study, an additional mid-feeder RL recloser relay will be placed at bus 828 to evaluate the coordination performance between RL relays. The mid-feeder recloser should respond only for faults in the second half of the feeder, and for these faults the substation recloser should operate after a delay. Autonomous micro-grid operation is not considered for this "proof of concept" study, although it could (and should) be an important direction for future studies. 
 
\begin{table}
	\begin{center}
		\caption{Difference Between OpenDSS and IEEE Solution}
		\label{pf_comparison}
		\begin{tabular}{|c||c||c||c|}
			\hline
			\label{tab_AS}
			\% Error & $V_a$ & $V_b$ & $V_c$ \\
			\hline
			Average & 0.179 & 0.240 & 0.023 \\
			\hline
			Maximum & 0.637 & 0.554 & 0.066 \\
			\hline
		\end{tabular}
		\vspace{-0.4cm}
	\end{center}
	
\end{table}


 Modifications to the  IEEE cases are done  when initializing each \emph{episode}  to simulate the real fluctuations of distribution grids. An episode is defined as a short simulation segment that contains a fault. A scenario is generated for each episode using a random combination of load and DER generation profile, fault parameter and fault location. The load and DER generation capacities are sampled from the COVID-EMDA+ dataset\cite{emda}, which has the real hourly renewable generation and load data for cities within each RTO region. In the beginning of each episode, a random hour is chosen from the year 2019, and the recorded load profile and PV capacity for Houston, Texas corresponding to that time is used to scale the load and PV generators. The maximum total installed capacity of PV is set to 30\% of the total load in the feeder and the locations are randomly scattered throughout all single-phase loads. The randomization of DER placement is only meant to provide singular experimental scenarios, although we are aware of the fact that the placement will have an impact on relay performance and would require more thorough analyses. For larger systems, techniques in \cite{DERdata} could potential be used to reduce the amount of computation power required.

In the middle of an episode, a random fault is added to the system. The fault will occur in a random line and phase(s), have a random impedance from 0.001 ohm to 20 ohm. All types of faults (SLG, LL, LLG, 3-phase) are possible. To match realistic scenarios in distribution lines, single phase faults have the highest chance to be selected and 3-phase faults have the lowest chance. The performance of the RL relays are evaluated by running a large number of random episodes. In the following demonstration, we run 5000 independent episodes for each type of scenario.
\vspace{-0.2cm}
\subsection{Overcurrent Protection}
 To set a baseline for comparison, a simple overcurrent recloser is placed at the substation and is configured to respond to faults in the distribution feeder. The settings of the overcurrent recloser relay is assumed to be twice the nominal current under the base case, in which the load capacities are the same as IEEE published numbers and the substation transformer is the only power supply for the feeder. A mid-feeder overcurrent recloser will also be placed at bus 828. The setting for the two overcurrent recloser relays used for comparison are recorded in Table \ref{oc_settings}. The fault detection and coordination of the overcurrent relays are tested under the basic IEEE 34 node feeder without considering any DER or load variation. The results in Table \ref{oc_baseline} shows the overcurrent relays are very reliable under the static environment. More specifically, the few times the overcurrent relays fail to detect faults are for single-phase faults in the 4.16 kV buses (888 and 890) with a relatively high fault impedance.

\begin{table}
	\begin{center}
		\caption{Settings for Baseline Overcurrent Relays}
		\label{oc_settings}
		\begin{tabular}{|c||c||c||c||c|}
			\hline
			Bus & Curve Type & Pickup Current(A) & Time Dial \\
			\hline
			800 & IEEE Very Inverse & 90  & 0.2 \\
			\hline
			828 & IEEE Very Inverse & 75  & 0.1 \\
			\hline
		\end{tabular}
	\end{center}
	
\end{table}

\begin{table}[t]
	\begin{center}
		\caption{Performance of Overcurrent Relays in Base Case}
		\label{oc_baseline}
		\begin{tabular}{|c||c||c|}
			\hline
			Type & Occurrences & Probability \\
			\hline
			False Alarm & 0 / 5000 & 0 \% \\
			\hline
			Fail to Detect & 21 / 5000 & 0.42 \% \\
			\hline
			Mis-Coordination & 13 / 5000 & 0.26 \% \\
			\hline
		\end{tabular}
	\end{center}
	\vspace{-0.2cm}
\end{table}

\section{Simulation Results}
\label{sec:simulations}

\subsection{Performance Metrics}
In this section we present and discuss  the performance of our Nested RL algorithm  for protective relays.  We compare the performance with conventional overcurrent relay protection strategy.  The performance is evaluated in  three aspects:

\noindent\textbf{Failure Rate}:  A relay failure happens when a relay fails to operate as it is expected to do. For each episode, we determine the optimal relay action from the type, time, and location of the fault, and compare it to the action taken by the  RL based relay.  We evaluate the percentage of the operation failures of the relays in four different scenarios: when there is a (i) fault in the local region, (ii) fault in the immediate downstream region, (iii) fault in a remote region, (iv) no fault in the network.  

\noindent\textbf{Robustness}: The load profiles in power distribution systems is a combined result from factors including renewable generation, load ramping, weather and social events. Moreover, both the total load capacity and renewable penetration are expected to grow consistently each year. Increase in the load capacity can cause a higher peak load and high renewable penetration can increase the variance of the load profile. It would be desirable if the protection system is robust against such changes to avoid the additional cost introduced by re-analyzing and re-programming the relays after deployment. We evaluate the performance of RL relays when the operating condition exceeds the nominal range.

\noindent\textbf{Response Time}: The response time of RL relay is defined as the  time difference between the inception of the fault and the relay decision and action. Response time is extremely critical in preventing hazards. For example,  it is preferred for the substation recloser to attempt clearing transient faults before any fuse in the feeder melts. This requires the recloser to have a fast fault detection time.  We compare the response time of the RL based relays with the conventional overcurrent relays.

\vspace{-0.3cm}
 
\subsection{Performance: Single Agent RL}


We first present the performance of our RL algorithm for a single recloser control. This is a special case of the proposed algorithm (with $n=1$).  We  train and test our  algorithm   in the context of substation recloser control in distribution feeders.  In particular, we consider a recloser located at the substation. The IEEE 34 bus feeder is used in this experiment.

 We have run the simulations with both the overcurrent protection and RL protection programmed in the same simulation setting. The same sequence current measurement from OpenDSS is provided to both the overcurrent and RL relay. The RL relay will remember each measurement value for a few steps and use the time window as input, while the overcurrent relay could be triggered by each incoming measurement snapshot. The simulation is run for 5000 randomly generated episodes and the operation of RL relay and overcurrent relay is logged and compared.

\begin{table}
	\begin{center}
		\caption{Failure Rate of Relays Under 30\% DER}
		\label{fail_rate_34_sq}
		\begin{tabular}{|c||c||c||c|}
			\hline
			\multicolumn{4}{|c|}{RL Based Relay} \\
			\hline
			\ Scenario & False Operation & Occurrences & Probability \\
			\hline
			No Fault & Trip & 0 / 5000 & 0.00 \% \\
			\hline
			Faulted & No Response & 16 / 5000 & 0.32 \% \\
			\hline
			\multicolumn{4}{|c|}{Overcurrent Relay} \\
			\hline
			\ Scenario & False Operation & Occurrences & Probability \\
			\hline
			No Fault & Trip & 0 / 5000 & 0.00 \% \\
			\hline
			Faulted & No Response & 773 / 5000 & 15.46 \% \\
			\hline
		\end{tabular}
	\end{center}
	\vspace{-0.3cm}
\end{table}



\begin{figure}[t]
	\centering
	\includegraphics[scale=0.4]{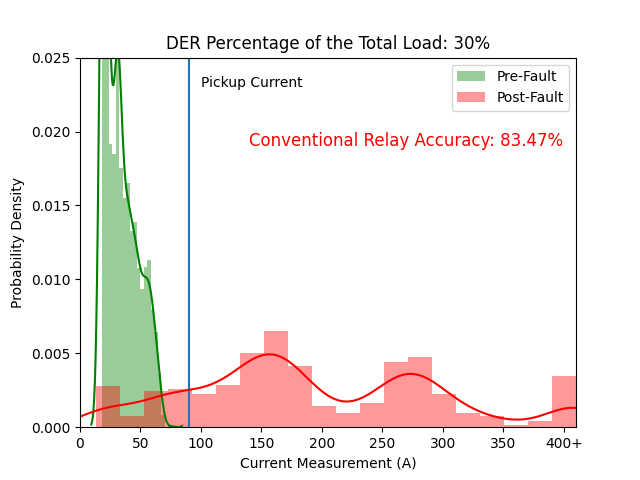}
	\caption{Pre-fault and Faulted Current Distribution at Bus 800}
	\label{fig:currdist}
	\vspace{-0.3cm}
\end{figure}

\begin{figure}[t]
	\centering
	\includegraphics[scale=0.4]{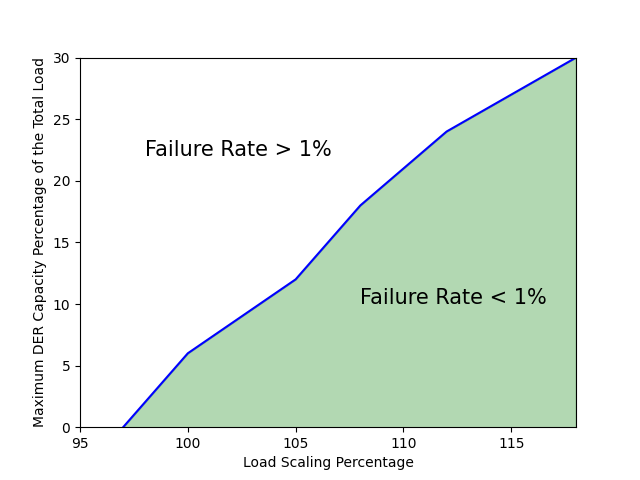}
	\caption{Failure Rate of Overcurrent Relay Under Different Load Scale and DER Penetration}
	\label{fig:loadvsder}
	\vspace{-0.3cm}
\end{figure}

Table \ref{fail_rate_34_sq} summarizes the {\bf failure rate} performance of both the RL relay and overcurrent relay in 34 bus test feeder.  The RL based relays are extremely accurate even under very high DER penetration levels. The fault current contribution from DER and fault impedance can, under many cases, reduce the magnitude of fault current measured at the substation (bus 800) considerably. As shown in Fig. \ref{fig:currdist}, the fault current magnitude can be very close tho the normal load current range for faults near the end of the feeder, high-impedance faults or faults in the two 4.16kV buses. Under these scenarios, a fixed pickup current can never completely separate the normal and fault condition because their distributions are overlapping. Fig. \ref{fig:loadvsder} presents a more direct illustration by plotting the line beyond which the failure rate of the overcurrent relay exceeds 1\% in the load scaler and DER capacity percentage space. The probability at each point is calculated from running 1000 random scenarios. 

To quantify the {\bf robustness} of RL based algorithm against peak load variations, the total load capacity the system is increased to up to 30\% more than the peak capacity used to generate the training data. In creating the validation data for robustness assessment, we focus on the robustness only when the system load is around the peak. For evaluating the robustness at 10\% higher load, the data is only selected when the system load is between 100\% and 110\% of the original capacity. Note that the model and policy of the RL relay remain unchanged, which means the data samples at the higher load are not used in training. The performance of RL relay under higher peak is shown in Table \ref{fail_rate_increased_peak}. 

\begin{table}
    \vspace{-0.2cm}
	\begin{center}
		\caption{Robustness Against Peak Load and DER Increase}
		\label{fail_rate_increased_peak}
		\begin{tabular}{|c||c||c||c|}
			\hline
			Peak Load  Increase& 10 \% & 20 \% & 30 \% \\
			\hline
			RL Failure Rate & 0.38 \% & 0.36 \% & 0.40 \%\\
			\hline
			Overcurrent Failure Rate & 6.5 \% & 7.4 \% & 9.8 \%\\
			\hhline{|=|=|=|=|}
			Peak DER Increase& 10 \% & 20 \% & 30\% \\
			\hline
			RL Failure Rate & 0.48 \% & 0.56 \% & 0.88 \%\\
			\hline
			Overcurrent Failure Rate & 18.5 \% & 19.7 \% & 22.9 \%\\
			\hline
		\end{tabular}
	\end{center}
	
\end{table}



Similarly, we also evaluate the robustness against potential increase in DER penetration. As the capacity of DERs in the distribution systems is expected to increase over time, it is desirable that the protection devices can reliably function without the need to re-configure their settings. In this experiment, the RL relays are trained using data created assuming an up to 30\% DER penetration as described in Sec. IV, B. The obtained policy is tested under scenarios where the DER penetration is increased above 30\%. The results are shown in the bottom half of Table \ref{fail_rate_increased_peak}. It can be seen that the RL relay is able to retain a good performance even when the DER penetration exceeds the amount that it is designed to operate on.

\begin{table}[t]
	\begin{center}
		\caption{Response Speed After Faults}
		\label{response_speed_sa}
		\begin{tabular}{|c||c||c||c||c|}
			\hline
			Delay & 1 Step & 2 Step & 3 Step & 4 Steps\\
			\hline
			Occurrences & 0 / 5000 & 4981 / 5000 & 17 / 5000 & 2 / 5000\\
			\hline
		\end{tabular}
	\end{center}
	\vspace{-0.3cm}
\end{table}

We also measure the {\bf response time} during the tests, quantified in terms of the number of simulation steps where each simulation step is 0.002 second. This step length is limited by the computation speed of the deep neural network model, which could be significantly improved with highly likely advances in hardware and software. The RL relays have shown a very small response time as listed in Table \ref{response_speed_sa}, the longest delay is 4 simulation steps which corresponds to 8 ms. Moreover, the fault detection time of RL relay is not explicitly correlated with fault current magnitude, and is much faster than the melting time curve of typical time-delay fuses under all scenarios. We note that, in practice however, the response time could be limited by the data acquisition rate of current measurements of instrument transformers. This fast response time also allows an ample time window for additional confidence check, during which successive flags can be used to reduce false-positives even further.

\vspace{-0.3cm}
 \subsection{Performance: Multi-Agent RL}


In distribution systems, it is common for large and long feeders to have additional protection in the middle of the feeder for additional security. For persisting faults that cannot be cleared by reclosing, the recloser needs to be locked open. In such cases, it is preferred for the closest protection device to operate to reduce the amount of load being disconnected and mitigate the damage. In the IEEE 34 bus case (Fig. 1), a mid-feeder recloser is added to the branch between bus 828 and bus 830. The mid-feeder recloser is expected to act immediately for all faults in the right half of the circuit, and remain closed for all faults between it and the substation. The substation recloser needs to provide backup for the mid-feeder recloser, taking a longer time delay for all faults pass the mid-feeder recloser and trip quickly for all faults between the substation and the mid-feeder recloser. Both the RL recloser relay control and the overcurrent relay control using inverse-time curves are implemented and compared.

\begin{figure}[t]	
	\captionsetup[subfigure]{justification=centering}
	\centering
	\includegraphics[width = 3in]{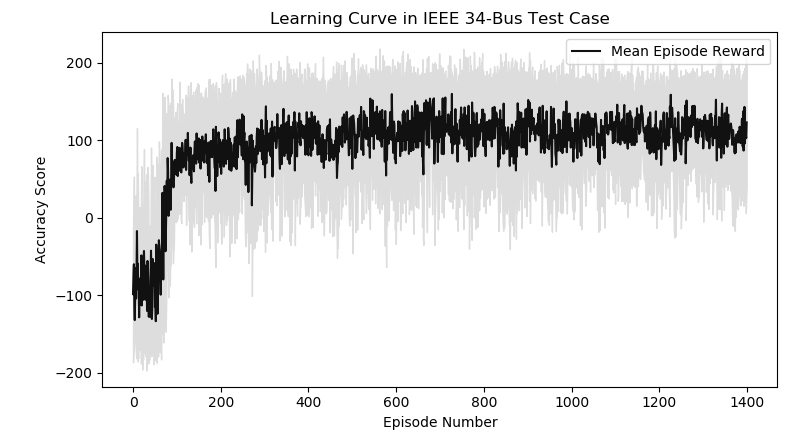}
	\caption{Learning Curve of Nested Multi-Agent RL}
	\label{fig_learningCurveMA}
\end{figure}

Our nested RL algorithm makes use of the radial structure of distribution grids. By this approach, if a relay need to provide backup for a downstream neighbor, it learns the optimal time delay before tripping the breaker for each possible fault scenario to accommodate the policy of its neighbors. For simulating the scenarios when a backup operation is needed during training, the substation recloser can only get a reward for tripping for faults after the mid-feeder recloser has attempted to trip. If the operation of the substation recloser is faster than the mid-feeder recloser, it will receive a penalty instead. According to our nested RL algorithm, the mid-feeder recloser is trained first with the substation recloser de-activated, and then the substation recloser is trained with the mid-feeder recloser put into action. We present the learning curve for the convergence of episodic reward of the substation recloser in Fig. \ref{fig_learningCurveMA} to demonstrate the good convergence even under multi-agent configuration. 

The {\bf failure rate} of the recloser pair is measured based on the action of both relays. An episode is considered successful only if both recloser take the correct control actions. The operation is tested in 5000 random episodes and the result is summarized in Table \ref{fail_rate_34_ma}.

\begin{table}[t]
	\begin{center}
		\caption{Failure Rate of Multi-Agent Relays}
		\label{fail_rate_34_ma}
		\begin{tabular}{|c||c||c|}
			\hline
			\ Scenario & Occurrences & Probability \\
			\hline
			\multicolumn{3}{|c|}{RL Based Relay} \\
			\hline
			False Alarm & 0 / 5000 & 0.00 \% \\
			\hline
			Fail to Detect & 19 / 5000 & 0.38 \% \\
			\hline
			Coordination Failure & 64 / 5000 & 1.28 \% \\
			\hline
			\multicolumn{3}{|c|}{Overcurrent Relay} \\
			\hline
			False Alarm  & 0 / 5000 & 0.00 \% \\
			\hline
			Fail to Detect & 696 / 5000 & 13.92 \% \\
			\hline
			Coordination Failure & 315 / 5000 & 6.30 \% \\
			\hline
		\end{tabular}
	\end{center}
	\vspace{-0.5cm}
\end{table}

{\bf Robustness} against increased peak load and DER capacity are conducted for the two-recloser pair similar to the single recloser scenario. A mis-operation of one recloser is recorded as failure for the entire episode. The results are listed in Table \ref{fail_rate_increased_peak_ma}. The impact of peak shift is slightly more evident than in previous single-relay cases due to the need for coordination and the performance of RL relays starts to deteriorate at around 15\% increased peak.

\begin{table}[t]
	\begin{center}
		\caption{Robustness Against Peak Increase: Multi-Agent}
		\label{fail_rate_increased_peak_ma}
		\begin{tabular}{|c||c||c||c|}
			\hline
			Peak Load Increase & 10 \% & 20 \% & 30  \%\\
			\hline
			RL Failure Rate & 0.62 \% & 0.69 \% & 0.83  \%\\
			\hline
			Overcurrent Failure Rate & 7.1 \% & 8.8 \% & 10.4 \%\\
			\hhline{|=|=|=|=|}
			Peak DER Increase& 10 \% & 20 \% & 30\% \\
			\hline
			RL Failure Rate & 1.02 \% & 1.16 \% & 1.30  \%\\
			\hline
			Overcurrent Failure Rate & 20.6 \% & 21.5 \% & 22.9 \%\\
			\hline
		\end{tabular}
	\end{center}
	
\end{table}

 The {\bf response time} for the both reclosers is recorded in Table \ref{response_speed}. It can be seen that the substation recloser responds faster to faults that are between the substation and the mid-feeder recloser. For faults in the right half of the circuit, the substation recloser provides a time window of roughly 3 time steps for the closer neighbor to operate first.
\begin{table}[t]
	\begin{center}
		\caption{Response Time}
		\label{response_speed}
		\begin{tabular}{|c||c||c||c||c|}
			\hline
			Delay & 1 Step & 2 Step & 3 Step & 4+ Steps\\
			\hline
			\makecell{Mid-feeder\\recloser} & 0/5000 & 4962/5000 & 28/5000 & 10/5000\\
			\hhline{|=|=|=|=|=|}
			Delay & 3- Step & 4 Step & 5 Step & 6+ Steps\\
			\hline
			\makecell{Substation\\recloser} & 2910/5000 & 345/5000 & 1591/5000 & 154/5000\\
			\hline
		\end{tabular}
		\vspace{-0.3cm}
	\end{center}
	
\end{table}

\section{Concluding Remarks}
\label{sec:conclusion}
This paper introduces and thoroughly tests a deep reinforcement learning based protective relay control strategy for the distribution grid with many DERs. It is shown that the proposed algorithm builds upon existing hardware and uses the same information available to today’s overcurrent protection yields much faster and more consistent performance. This algorithm can be easily applied in both a standalone relay and a network of coordinating relays. The trained RL relays can accurately detect faults under situations including high fault impedance, presence of distributed generation and volatile load profile, where the performance of traditional overcurrent protection deteriorates heavily. The RL relays are robust against unexpected changes in operating conditions of the distribution grid at the time of planning, reducing the need to re-train the relays after deployments. The fast response speed provides ample time for coordinating with fuses and other relays.

The proposed deep RL relays will be easy to implement with the currently available distribution infrastructure. A particularly attractive feature is that the proposed algorithm for relays can operate in a completely decentralized manner without any communication. This communication-free setting is not only easy to implement for currently available distribution grid infrastructure, but also less vulnerable to potential cyber-attacks. The input to the RL relays are the same as traditional relays so the instrument transformers can be retained during deployment. The training process does not require human intervention since the production of training data and computation of optimal control policy can be fully automated. The weights of the DQN obtained during training can be saved into a general-purpose micro-controller or potentially a more optimized machine learning chip.

In the future, we plan to provide a theoretical guarantee for the convergence of our sequential RL algorithm. We will conduct a thorough and careful investigation of the operation of RL relays under various realistic scenarios by running year-long simulations under variety of stochastic variation of the operational and fault parameters of networks. We will explore potential performance or robustness improvement by using more inputs parameters such as voltage, frequency or apparent impedance. We are working with time domain simulators for more detailed training data generation and fault study with the most realistic models for components including control loops and electromagnetic transients. We will also investigate the possibility of hardware prototyping and Hardware-in-the-Loop test with Real-Time Digital Simulator (RTDS).

\ifCLASSOPTIONcaptionsoff
  \newpage
\fi

%








\end{document}